\documentclass[%
 aip,
 jcp,
 amsmath,amssymb,
 reprint,%
]{revtex4-1}

\usepackage{graphicx}
\usepackage{dcolumn}
\usepackage{bm}

\usepackage[utf8]{inputenc}
\usepackage[T1]{fontenc}
\usepackage{mathptmx}
\usepackage{etoolbox}
\usepackage{comment}
\usepackage{tikz}
\usepackage{hyperref}
\usepackage{dsfont}
\usepackage{array}
\usepackage{relsize}
\usepackage{yfonts}
\usepackage{cases}
\usepackage{adjustbox}
\usepackage{url}

\makeatletter
\def\@email#1#2{%
 \endgroup
 \patchcmd{\titleblock@produce}
  {\frontmatter@RRAPformat}
  {\frontmatter@RRAPformat{\produce@RRAP{*#1\href{mailto:#2}{#2}}}\frontmatter@RRAPformat}
  {}{}
}%
\makeatother
\begin{document}

\preprint{AIP/123-QED}

\title{Molecular Axis Distribution Moments in Ultrafast Transient Absorption Spectroscopy: A Path Towards Ultrafast Quantum State Tomography}



\author{Shashank Kumar}
    \email{kumar414@purdue.edu}
    \affiliation{Department of Physics and Astronomy, Purdue University, West Lafayette, 47907, Indiana, USA}%

\author{Eric Liu}%
    \affiliation{Department of Physics and Astronomy, Purdue University, West Lafayette, 47907, Indiana, USA}%

\author{Liang Z. Tan}
    \affiliation{Molecular Foundry, Lawrence Berkeley National Laboratory, Berkeley, California 94720, USA}

\author{Varun Makhija}
    \email{vmakhija@umw.edu}
    \affiliation{Department of Chemistry and Physics, University of Mary Washington, Fredericksburg, Virginia 22401, USA}

\author{Niranjan Shivaram}%
    \email{niranjan@purdue.edu}
    \affiliation{Department of Physics and Astronomy, Purdue University, West Lafayette, 47907, Indiana, USA}%
    \affiliation{Purdue Quantum Science and Engineering Institute, Purdue University, West
    Lafayette, 47907, Indiana, USA.}

\date{\today}

\begin{abstract}
In ultrafast experiments with gas phase molecules, the alignment of the molecular axis relative to the polarization of the interacting laser pulses plays a crucial role in determining the dynamics following this light-matter interaction. The molecular axis distribution is influenced by the interacting pulses and is intrinsically linked to the electronic coherences of the excited molecules. However, in typical theoretical calculations of such interactions, the signal is either calculated for a single molecule in the molecular frame or averaged over all possible molecular orientations to compare with the experiment. This averaging leads to the loss of information about anisotropy in the molecular-axis distribution, which could significantly affect the measured experimental signal. Here, we calculate the laboratory frame transient electronic first-order polarization ($P^{(1)}$) spectra in terms of separated molecular frame and laboratory frame quantities. The laboratory frame polarizations are compared with orientation-averaged Quantum Master Equation (QME) calculations, demonstrating that orientation-averaging captures only the isotropic contributions. We show that our formalism allows us to also evaluate the anisotropic contributions to the spectrum. Finally, we discuss the application of this approach to achieve ultrafast quantum state tomography using transient absorption spectroscopy and field observables in nonlinear spectroscopy.
\end{abstract}

\maketitle


\section{Introduction}
Ultrafast laser pulses are extensively used to investigate electronic coherences in molecules, as they can generate broadband electronic wavepackets through a superposition of excited states within the molecule \cite{Goulielmakis2010, calegari2014, 
 Makhija_2020}. Previous studies have explored electronic coherence using various techniques, including high-harmonic generation (HHG) \cite{PhysRevA.91.023421, PhysRevLett.111.243005}, photoelectron spectroscopy \cite{bennett2016nonadiabatic, Makhija_2020, Cavaletto2022, PhysRevLett.131.193001, kaufman2024}, XUV Transient Absorption Spectroscopy ~\cite{kobayashi2019,chang2020} and ultrafast X-ray scattering \cite{simmermacher2019, PhysRevLett.115.193003, Kowalewski2017}. Since rotational motion significantly influences the dynamics of electronic coherences \cite{PhysRevA.95.033425, PhysRevLett.118.083001,Makhija2022}, it is often necessary to transform laboratory-frame (LF) observables to the molecular-frame (MF). Such transformations have been extensively studied for photoelectron spectroscopy where photoelectron angular distributions (PADs) are experimentally accessible \cite{underwood2000,reid2003, hockett20072, Gregory2021}. One of the key advantages of LF-to-MF transformations is their ability to reconstruct a molecule's electronic density matrices, a process called quantum tomography\cite{dunn1995, dariano2003, skovsen2003, mouritzen2006, PhysRevLett.131.193001}.

Here, we develop an LF-MF transformation for the case of transient absorption spectroscopy (TAS), which is a special case of third-order nonlinear spectroscopy. In most ab initio calculations of TAS for molecules, the output spectra are averaged over all molecular orientations \cite{Golubev2021, Rott2021}. However, this averaging does not capture the anisotropy of the molecular axis distribution created during light-matter interaction leading to a loss of valuable information.We demonstrate that the contribution from an anisotropic molecular axis distribution to the experimental signal can be reconstructed using measured linear and nonlinear optical observables. In Sec~\ref{subsection: LF_P1}, the first-order laboratory frame electronic polarization (LF) is derived using perturbation theory, expressed in terms of molecular frame (MF) dipole moments. Under the assumption that the populations and coherences do not change over the duration of the probe pulse, the first-order polarization can be expanded in terms of molecular axis distribution moments (MADM)\cite{Makhija2022}, which contain information on the anisotropy of the axis distribution of molecules in the laboratory frame while the corresponding coefficients encode the information about the molecular frame. In Sec~\ref{subsection: ORRCS}, this transformation is used to derive molecular-frame polarization in the orientation resolution through the rotational coherence spectroscopy (ORRCS) technique \cite{makhija2016, PhysRevA.96.023424}. ORRCS can extract the MF polarization from the LF measured signal by calculating the MADMs using the calculated alignment distribution of the molecules for the experimental parameters. To demonstrate proof of concept, a detailed comparison of the LF first-order polarization calculation with an orientation-averaged Quantum Master Equation (QME) simulation of the transient absorption spectroscopy experiment is provided in Sec~\ref{section: TDSE}. Additionally, in Sec.~\ref{Quantum_tomography}, we use this formalism to develop a scheme for transient absorption spectroscopy through which all the relevant MADMs can be measured for an electronic wavepacket in ammonia, paving the path to Quantum State Tomography.

\section{Theory}
\subsection{Precursors}

The most general non-relativistic Hamiltonian of a molecule includes both electronic and nuclear components. For a non-adiabatic system, the complete wavefunction corresponding to such a  Hamiltonian can be expanded in a complete orthonormal set of basis function\cite{Stolow2008, Stanke2017},
\begin{eqnarray}
    \psi(\bm{r},\bm{R}) = \sum_{n} \chi_n(\bm{R})\phi_n(\bm{r},\bm{R})
\end{eqnarray}
where, $\chi_n(\bm{R})$ is the nuclear wavefunction and $\phi_n(\bm{r},\bm{R})$ is the electronic wavefunction for the electronic state $\lvert n\rangle$. However, the Born-Oppenheimer (BO) approximation assumes that there is no coupling between nuclear and electronic states, as the much lighter mass of electrons allows them to respond almost instantaneously to the motion of nuclei. Therefore, within this approximation, the electronic wavefunction $\phi_n(\bm{r},\bm{R})$ is a function of the electronic coordinates $\bm{r}$ and the nuclear coordinates $\bm{R}$ appear as parameters. The nuclear wavefunction $\chi(\bm{R})$ can be described as a combination of the vibrational wavefunction of the nuclei of the molecules and the rotational wavefunction of the molecule. Vibrational motion can be approximated using an effective harmonic oscillator \cite{Roy2009}, whereas rotational dynamics has been extensively studied by solving the rigid rotor Hamiltonian\cite{Zare1988}. The centrifugal and Coriolis couplings are neglected, which allows the vibrational and rotational wavefunction to be separated\cite{Stolow2008, Roy2009, Zare1988}. Therefore, a wavepacket in the laboratory frame (LF) can be written as a superposition of decoupled rovibronic states in the BO basis as\cite{Stolow2008, Makhija2022},
\begin{align}
\label{gen-wavefunction}
    \lvert \Psi(t)\rangle = \sum_{\substack{J_\alpha,M_\alpha,\tau_\alpha\\v_\alpha,\alpha}}C_{\substack{J_\alpha,M_\alpha,\tau_\alpha\\v_\alpha,\alpha}}(t)e^{-i\mathcal{E}_{J_\alpha, M_\alpha, \tau_\alpha}^\alpha t} \lvert J_\alpha M_\alpha \tau_\alpha \rangle \lvert v_\alpha\rangle \lvert \alpha \rangle
\end{align}
where, $\lvert J_\alpha M_\alpha \tau_\alpha \rangle$ are the asymetric top eigenstates~\cite{Zare1988}. $J_\alpha, M_\alpha$ are the total angular momentum quantum number and its projection to z-axis of the lab frame, respectively. $\tau_\alpha$ is not an angular momentum quantum number but is used as a label to keep track of the $2J+1$ different ($J_\alpha, M_\alpha$) rotational energy states. A similar equation can also be written in the symmetric top basis\cite{Zare1988}. $\lvert v_\alpha\rangle, \lvert\alpha\rangle$ are the vibrational and electronic states. The $\mathcal{E}_{J_\alpha, M_\alpha, \tau_\alpha}^\alpha = \mathcal{E}_{\alpha, v_\alpha} + \mathcal{E}_{J_\alpha, M_\alpha, \tau_\alpha}$ are the eigen energies of the BO states which is the sum of vibronic and rotational energies. The corresponding rotational, vibrational, and electronic wavefunctions respectively are\cite{Stolow2008},
\begin{subequations}
\label{vibronic-wavefunction}
    \begin{align}
        \langle \bm{\Omega}\lvert J_\alpha M_\alpha \tau_\alpha \rangle &= \psi_{J_\alpha M_\alpha \tau_\alpha}(\bm{\Omega})\\
        \langle \bm{R}\lvert v_\alpha \rangle &= \psi_{v_\alpha}(\bm{R})\\
        \langle \bm{r},\bm{R}\lvert \alpha \rangle &= \phi_{\alpha}(\bm{r,}\bm{R})
    \end{align}
\end{subequations}
where $\bm{\Omega}=\{\phi,\theta,\chi\}$ are the Euler angles when rotated from a Lab-frame (LF) to the Molecular-frame (MF)~\cite{Zare1988}.

\subsection{First order polarization in the Laboratory frame}\label{subsection: LF_P1}
The total time-dependent polarization can be written as,
\begin{equation}
    \bm{P}(t) = \langle \Psi(t) \lvert \bm{\mu} \lvert \Psi(t) \rangle
\end{equation}
Using time-dependent perturbation theory, the first-order polarization is written as,
\begin{align}
\label{P_1}
    P^{(1)}(t) = & i\langle \Psi(0)\lvert U^{\dagger}(t) (\bm{\mu}\cdot \hat{\epsilon}) \notag \\
    &\int_{-\infty}^t dt' U(t-t') (\bm{\mu}\cdot\hat{\epsilon}) E(t') U(t') \lvert \Psi(0) \rangle
    + \text{c.c.}
\end{align}
Here, $U(t)$ is the evolution operator and defined as,
\begin{align}
\label{BO_basis}
    &U(t)\lvert \Psi(0)\rangle = \lvert \Psi(t)\rangle \notag\\
    &= \sum_{n_\alpha, M_\alpha} C_{n_\alpha, M_\alpha}(t) e^{-i\mathcal{E}_{J_\alpha, K_\alpha, M_\alpha}^\alpha t} \lvert J_\alpha K_\alpha M_\alpha \rangle \lvert v_\alpha\rangle \lvert \alpha \rangle
\end{align}
where $n_\alpha = J_\alpha,K_\alpha, v_\alpha,\alpha$ is used for brevity. The most general state is written as a superposition of eigenstates of the symmetric top Hamiltonian in contrast to Eq.~\ref{gen-wavefunction}. This representation introduces a new angular momentum quantum number $K_\alpha$ which is the projection of total angular momentum onto the z axis of the molecular frame. This wavepacket can be generated using a broadband pump laser pulse, enabling the population of multiple excited states\cite{Stolow2008}.

We assume that a probe pulse interacts with a pump-excited wavepacket and creates a first-order electronic polarization. Substituting Eq~\ref{BO_basis} into Eq~\ref{P_1} and performing the Fourier transform yields the first-order polarization spectrum
under the assumption that the coefficients $C_{n_\alpha, M_\alpha}(t)$ do not vary over the duration of the probe pulse. This implies that the population/coherence created by the pump pulse is unaffected by the probe pulse\cite{Stolow2008}. Consequently, the coefficients $C_{n_\alpha, M_\alpha}(t)$ are parametrized by the time delay between the pump and probe pulses, $\tau$. The first-order polarization spectrum in the LF is then expressed as \cite{Golubev2021},
\begin{widetext}
\begin{align}
\label{P1_LF}
    \Tilde{P}^{(1)}(\omega; \tau) = &\sum_{\substack{n_\alpha, M_\alpha \\ n_{\beta}, M_{\beta}}}C_{n_\alpha, M_\alpha}(\tau) C_{n_\beta, M_\beta}^*(\tau) \sum_{n_\gamma, M_\gamma}\langle n_\beta M_\beta\lvert \bm{\mu} \cdot \hat{\epsilon} \lvert n_\gamma M_\gamma \rangle \langle n_\gamma M_\gamma \lvert \bm{\mu} \cdot \hat{\epsilon} \lvert n_\alpha M_\alpha \rangle\notag \\
    & \times \underbrace{\Tilde{E}(\omega - \mathcal{E}_{n_\beta, M_\beta} + \mathcal{E}_{n_\alpha, M_\alpha}) \left(\frac{1}{\mathcal{E}_{n_\gamma, M_\gamma} - \mathcal{E}_{n_\beta, M_\beta} - \omega - i\Gamma_{\gamma \beta}} + \frac{1}{\mathcal{E}_{n_\gamma, M_\gamma}^* - \mathcal{E}_{n_\alpha, M_\alpha} + \omega + i\Gamma_{\gamma\alpha}}\right)}_{=E'\left(\omega,\mathcal{E}_{J_\beta, K_\beta, M_\beta}^\beta, \mathcal{E}_{J_\alpha, K_\alpha, M_\alpha}^\alpha, \mathcal{E}_{J_\gamma, K_\gamma, M_\gamma}^\gamma \right)}
\end{align}
\end{widetext}
Here, $\Tilde{E}(\omega)$ is the spectrum of the probe pulse. The decay terms $\Gamma_{\gamma\alpha},\Gamma_{\gamma\beta}$ have been added phenomenologically. 

The LF dipole matrix elements $\langle n_\beta M_\beta\lvert \bm{\mu} \cdot \hat{\epsilon} \lvert n_\gamma M_\gamma \rangle$ can be transformed into the MF  by rotating its spherical components using the appropriate Euler angles (see supplementray material for full derivation).
\begin{align}
\label{LF_dm}
    \langle n_\beta M_\beta \lvert \bm{\mu} \cdot \hat{\epsilon} \lvert n_\gamma M_\gamma \rangle = \sum_{p'=-1}^{1}\sum_{q'=-1}^{1} (-1)^{p'}  e_{-p'} \mathcal{D}_{\beta \gamma}(q')\nonumber\\
    \times \int d\bm{\Omega} \frac{\sqrt{[J_\gamma, J_\beta]}}{8\pi^2} D_{p'q'}^{1*}(\bm{\Omega}) D_{M_\beta K_\beta}^{J_\alpha}(\bm{\Omega}) D_{M_\gamma K_\gamma}^{J_\gamma *}(\bm{\Omega})
\end{align}

where, $D^J_{MK}(\bm{\Omega})$ is the Wigner-D matrix. The notation $[X,Y,\cdots] = (2X+1)(2Y+1) \cdots$ is used for brevity. The $\mathcal{D}_{\beta \gamma}(q^{\prime})$ is the dipole matrix element in the MF between vibronic state $\gamma$ and $\alpha$ and defined by,
\begin{equation}
    \mathcal{D}_{\gamma \alpha}(q) = \int d\bm{R} \psi_{v_\gamma}^*(\bm{R}) \left[\int d\bm{r} \phi^*_\gamma(\bm{r},\bm{R}) d_q \phi_\alpha(\bm{r},\bm{R})\right] \psi_{v_\alpha}(\bm{R})
\end{equation}
The $\psi_{v_\alpha}, \phi_\alpha$ are the vibrational and electronic wavefunction defined in Eq~\ref{vibronic-wavefunction}.

Using the properties of the Wigner-D matrix (see supplementary material), the LF first-order polarization can be alternatively written in terms of Wigner-3j symbols,
\begin{widetext}
\begin{align}
\label{p_3j}
    \Tilde{P}^{(1)}(\omega; \tau) = &\sum_{\substack{n_\alpha, M_\alpha \\ n_{\beta}, M_{\beta}}} \sum_{n_\gamma, M_\gamma}\sum_{p,p'}\sum_{q, q'}(-1)^{p+p'}  e_{-p} e_{-p'} \mathcal{D}_{\beta \gamma}(q')  \mathcal{D}_{\gamma \alpha}(q) \sqrt{[J_\alpha, J_{\gamma}^2, J_\beta]} E'\left(\omega,\mathcal{E}_{J_\beta, K_\beta, M_\beta}^\beta, \mathcal{E}_{J_\alpha, K_\alpha, M_\alpha}^\alpha, \mathcal{E}_{J_\gamma, K_\gamma, M_\gamma}^\gamma\right) \notag \\
    & \times (-1)^{p-q} (-1)^{M_\alpha - K_\alpha} (-1)^{M_\beta - K_\beta} C_{n_\alpha, M_\alpha}(\tau) C_{n_\beta, M_\beta}^*(\tau) \notag\\
    & \times\begin{pmatrix}
        J_{\gamma} & J_\alpha & 1\\
        M_{\gamma} & -M_\alpha & -p
    \end{pmatrix}
    \begin{pmatrix}
        J_{\gamma} & J_\alpha & 1\\
        K_{\gamma} & -K_\alpha & -q
    \end{pmatrix}
    \begin{pmatrix}
        J_{\gamma} & J_\beta & 1\\
        M_{\gamma} & -M_\beta & p'
    \end{pmatrix}
    \begin{pmatrix}
        J_{\gamma} & J_\beta & 1\\
        K_{\gamma} & -K_\beta & q'
    \end{pmatrix}
\end{align}
\end{widetext}
To further simplify the expression, it is assumed that the rotational eigenstates are not resolved because the bandwidth of the probe pulse is much larger than the spacing between the rotational states. This helps in taking the summation over $J_\gamma, K_\gamma$ and $M_\gamma$ (i.e. rotational eigenstates) since the total energy $\mathcal{E}_{J_\gamma, K_\gamma, M_\gamma}^\gamma \approx \Bar{\mathcal{E}}_{\gamma} $ can be approximated with the energy averaged over all the rotational states\cite{Stolow2008, Zare1988}. Using the angular momentum mixing properties of the Wigner-3j symbol and the orthonormality condition of the Wigner-6j symbol, the first-order LF polarization can be further simplified to,
\begin{widetext}
\begin{align}
\label{P_w_t}
    P^{(1)}(\omega; \tau)&=\sum_K\sum_{p,p'}\sum_{q, q'}(-1)^{p+p'}  e_{-p} e_{-p'} 8\pi^2 \sum_{\alpha, v_\alpha}\sum_{\beta, v_\beta}\sum_{\gamma, v_\gamma}\mathcal{D}_{\beta \gamma}(q') \mathcal{D}_{\gamma \alpha}(q) E'(\omega,\Bar{\mathcal{E}}_\beta, \Bar{\mathcal{E}}_\alpha, \Bar{\mathcal{E}}_\gamma)
    \begin{pmatrix}
        1 & 1 & K\\
        -p' & -p & Q
    \end{pmatrix}
    \begin{pmatrix}
        1 & 1 & K\\
        -q' & -q & S
    \end{pmatrix}\notag\\
    &\times \sum_{J_\alpha K_\alpha M_\alpha} \sum_{J_\beta K_\beta M_\beta} \rho(n_\beta M_\beta; n_\alpha M_\alpha; \tau) \frac{\sqrt{[J_\alpha, J_\beta]}}{8\pi^2} (-1)^{M_\alpha - K_\alpha} (2K+1)
    \begin{pmatrix}
        J_\alpha & J_\beta & K\\
        -M_\alpha & M_\beta & Q
    \end{pmatrix}
    \begin{pmatrix}
        J_\alpha & J_\beta & K\\
        -K_\alpha & K_\beta & S
    \end{pmatrix}
\end{align}    
\end{widetext}
Here all the angular momentum quantum numbers are integers since we are excluding spin angular momentum coupling. In the last step, we identify $\rho(n_\beta M_\beta; n_\alpha M_\alpha; \tau) = C_{n_\alpha, M_\alpha}(\tau) C_{n_\beta, M_\beta}^*(\tau)$ as the LF density matrix elements in the BO basis.

We separate the time delay and frequency dependence of Eq~\ref{P_w_t} writing it as,
\begin{eqnarray}
    \label{P1_MADM}
    P^{(1)}(\omega; \tau) = \sum_{\alpha, v_\alpha}\sum_{\beta, v_\beta} \sum_{K,Q,S}C_{KQS}(\omega, \alpha, \beta, v_\alpha, v_\beta)\nonumber\\
    \times A_{QS}^K(\alpha, \beta, v_\alpha, v_\beta; \tau)
\end{eqnarray}
Here, the summations over $Q, S$ are redundant because of the property of the 3j symbol $-M_\alpha+M_\beta+Q=0$ and $-K_\alpha+K_\beta+S=0$ their values are fixed (but for the sake of generality they are kept in the expression). We identify the $A_{QS}^K(\alpha, \beta, v_\alpha, v_\beta; \tau)$ as the Molecular Axis Distribution Moments (MADMs) previously identified in the measurement of PADs\cite{Stolow2008, Makhija2022, Makhija_2020}. These MADMs contain LF time-dependent information about molecular orientation anisotropy after the pump pulse creates the coherence between the vibronic states $\lvert\alpha,v_\alpha\rangle$, and $\lvert\beta,v_\beta\rangle$. 

The coefficients $C_{KQS}(\omega, \alpha, \beta, v_\alpha, v_\beta)$  are given by,
\begin{align}
\label{C1_LF}
    C_{KQS}(\omega, \alpha, \beta, v_\alpha, v_\beta) = \sum_{p,p'}\sum_{q, q'}(-1)^{p+p'} 8\pi^2  e_{-p} e_{-p'} \notag\\
    \times \sum_{\gamma, v_\gamma}\mathcal{D}_{\beta \gamma}(q') \mathcal{D}_{\gamma \alpha}(q) E'(\omega,\Bar{\mathcal{E}}_\beta, \Bar{\mathcal{E}}_\alpha, \Bar{\mathcal{E}}_\gamma) \notag\\
    \times \begin{pmatrix}
        1 & 1 & K\\
        -p' & -p & Q
    \end{pmatrix}
    \begin{pmatrix}
        1 & 1 & K\\
        -q' & -q & S
    \end{pmatrix}
\end{align}

The above derivation shows that the signal of a pump-probe spectroscopy can be separated into two terms, $A_{QS}^K(\alpha, \beta, v_\alpha, v_\beta; \tau)$ which are the delay-dependent MADMs that describe the axis distribution of the molecule in the superposition of states ensemble created by the pump pulse and the frequency-dependent coefficients $C_{KQS}(\omega, \alpha, \beta, v_\alpha, v_\beta)$ which contain information about the excited states vibronic transitions due to the probe pulse and, importantly, the MF transition dipole moments. The expression Eq.~\ref{P1_MADM} thus transparently relates the first-order polarization to populations and coherences determining the molecular dynamics (the MADMs $A^K_{QS}$) and molecular frame quantities contributing to the probe step (the coefficients $C^K_{QS}$).

\subsubsection{Selection rules}
\label{subsubsection: Selection_rules}
Equation~\ref{P1_MADM} represents the most general form of the first-order polarization in the laboratory frame (LF). However, most angular momentum contributions become zero, depending on the molecular properties and the polarization of the incoming pulses. In this study, we focus on a system of linear molecules and linearly polarized light. For linear molecules, the molecular axis can be aligned with the z axis, and for linearly polarized light, the z axis of the laboratory frame can be aligned with the polarization direction. The Euler angles can then be chosen as $\Omega = \{\theta, \phi\}$ with $p,p'=0$ being the only non-zero component in Eq. \ref{p_3j} (see also Eq. S1 of the supplementary material). Under these conditions, the angular momentum quantum numbers required to completely describe the MADM's molecular orientation-dependent anisotropy are $Q=S=0$ \cite{Stolow2008, Zare1988}.

Starting from the rotational ground state of $|0,0,0\rangle$, the pump pulse excitation process, involving a single photon, leads to excited states with a total angular momentum of $J_\beta, J_\alpha = 1$\cite{Makhija2022}. This yields allowed values of quantum number $K=0,1,2$ using the properties of the Wigner 3j symbol in Eq.~\ref{P_w_t}. However, since the values of $p,p', Q = 0$ in Eq~\ref{C1_LF}, $K+2$ has to be an even number, which means that $K$ must be even. Therefore, for a system of linear molecules that interact with linearly polarized light, the allowed values of $K=0,2$ with $S = 0$ and $Q = 0$.

\subsection{Orientation Resolution through Rotational Coherence Spectroscopy (ORRCS)}
\label{subsection: ORRCS}
ORRCS is a technique used to extract MF information from the LF-measured signal. In previous studies, this method has been applied to photoionization signals \cite{makhija2016, PhysRevA.96.023424} and high-harmonic generation yield data\cite{Kanai2005}. In recent work\cite{Pandey24}, the ORRCS method was used with degenerate four-wave mixing (DFWM) signals to extract the MF temporal chirp of the output signal for linear molecules. Regardless of the observable, the LF-measured signal is consistently assumed to be a convolution of a response function $\mathcal{R}(\bm{\Omega})$ with the delay-dependent molecular axis distribution \cite{PhysRevA.96.023424},
\begin{eqnarray}
    \langle\mathcal{O}\rangle (\tau) = \int d\bm{\Omega}\rho(\bm{\Omega}, \tau) \mathcal{R}(\bm{\Omega})
\end{eqnarray}
where, $\bm{\Omega}$ is the Euler angles connecting LF to MF. For previous studies using the photoionization yield, $\mathcal{O}=S$\cite{makhija2016} and for  DFWM, $\mathcal{O}=a_2$, where $a_2$ is the temporal chirp coefficient extracted from the temporal phase\cite{Pandey24}.

This section provides a mathematically rigorous derivation for the LF first-order polarization so that an assumption of a response function is unnecessary for this observable. The MADMs are rewritten in the basis of Euler angles as \cite{Makhija2022},
\begin{align}
\label{euler_basis}
    A_{QS}^K(n, n'; \tau) = \frac{2K+1}{8\pi^2} \int d\bm{\Omega} D_{QS}^K(\bm{\Omega}) \rho_{n,n'}(\bm{\Omega}, \tau)
\end{align}
where $\rho_{n,n'}(\bm{\Omega}, t)$ is the time-dependent LF density matrix elements. Using this relation in Eq~\ref{P1_MADM},
\begin{align}
\label{gen_orrcs}
    P^{(1)}(\omega; \tau) = \sum_{\alpha, v_\alpha}\sum_{\beta, v_\beta} \sum_{K,Q,S} C_{KQS}(\omega, \alpha,v_\alpha, \beta,v_\beta) \notag\\
    \times \frac{2K+1}{8\pi^2} \int d\bm{\Omega} D_{QS}^K(\bm{\Omega}) \rho_{\alpha,v_\alpha,\beta,v_\beta}(\bm{\Omega}, \tau)
\end{align}
Assuming that the photon energy of the pump pulse is small and non-resonant such that the molecule remains in the vibronic ground state, the LF-density matrix $\rho_{\alpha,v_\alpha,\beta,v_\beta} = \rho$ becomes delay-dependent molecular axis distribution. Then,
\begin{align}
\label{P_orrcs_LF}
    P^{(1)}(\omega; \tau) = \sum_{K,Q,S} c_{KQS}(\omega)\langle D_{QS}^K\rangle(\tau)
\end{align}
where, $c_{KQS}(\omega) = \frac{2K+1}{8\pi^2}\cdot C_{KQS}(\omega)$ and the expectation value of the Wigner D-matrix is taken over a rotational wavepacket. A similar expression for the first-order MF signal can also be written as the expansion of Wigner-D matrices (see the supplementary material for full derivation),
\begin{align}
\label{P_orrcs_MF}
     P^{(1)}_{MF}(\omega; \Omega) = \sum_{K,Q,S} c^{MF}_{KQS}(\omega)D_{QS}^K(\bm{\Omega})
\end{align}

In an experiment, a low-photon-energy alignment pulse can create a rotational wavepacket in an ensemble of molecules. Then, a resonant probe pulse can induce a transition from the vibronic ground state to an excited state. The output pulse emitted by the time-dependent polarization is then measured as a function of the time delay ($\tau$) between the alignment and the probe pulse. Using the asymmetric rotor Hamiltonian, the rotational wavepacket can be calculated to find $\langle D_{QS}^K\rangle(\tau)$\cite{PhysRevA.81.043425}. The numerically calculated $\langle D_{QS}^K\rangle(\tau)$ can be fitted to the experimentally measured LF polarization to extract the coefficients $c_{KQS}(\omega)$ from Eq~\ref{P_orrcs_LF}. This coefficient represents the probe interaction involving excitation from $\lvert \alpha, v_\alpha\rangle$ to all possible $\lvert \gamma, v_\gamma\rangle \rightarrow \lvert \beta, v_\beta\rangle$. In the above mentioned experiment, $\lvert \alpha, v_\alpha\rangle, \lvert \beta, v_\beta\rangle = \lvert 0,0\rangle$ because the molecule is in the vibronic ground state. If the excited state energies $\mathcal{E}_\gamma$ are known, then the coefficient can be determined by fitting to data using Eq~\ref{C1_LF} to extract the Molecular Frame Transition Dipole Moments $\mathcal{D}_{\beta \gamma}(q'), \mathcal{D}_{\gamma \alpha}(q)$. For a system of linear molecules probed by linearly polarized light, there are two equations for each excited state $\lvert \gamma, v_\gamma\rangle$, corresponding to the coefficients $C_{000}(\omega), C_{200}(\omega)$ (see Sec~\ref{subsubsection: Selection_rules}). Thus, with two equations and two unknowns ($\mu_x=\mu_y, \mu_z$), it is possible to uniquely determine the transition dipole moments in the MF. From there it is straightforward to construct the MF polarization using Eq~\ref{P_orrcs_MF}.

\section{Comparison with Simulation}
\label{section: TDSE}

\subsection{Transient Absorption spectroscopy}

\begin{figure}[ht]
    \centering
    \includegraphics[width=\linewidth]{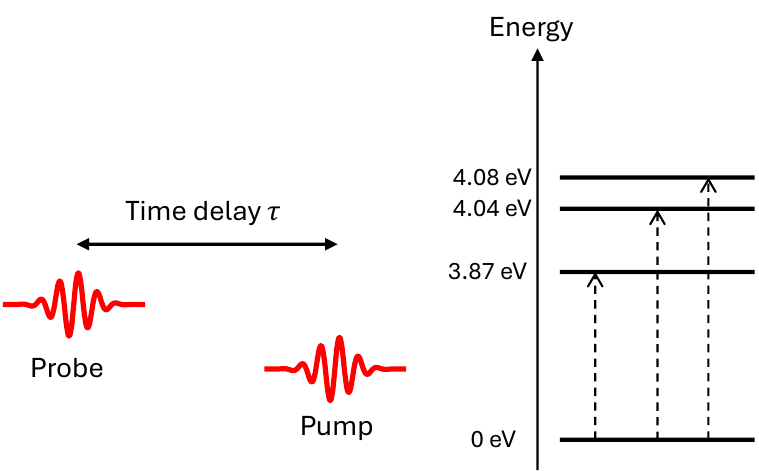}
    \caption{The schematic illustrates a transient absorption experiment involving a linear molecule. The dashed arrows in the energy level diagram correspond to the dipole-allowed transitions between the states. The pump and probe pulses have photon energies of $4\text{eV}$ with a time delay of $\tau$ fs between them. The pump pulse has a large enough bandwidth to populate the excited states coherently, and the measurement is performed along the direction of the probe pulse to determine the absorption spectrum.}
    \label{fig:ene_diag}
\end{figure}

A detailed QME simulation of a transient absorption experiment is performed on a fictitious linear molecule whose energy level diagram is shown in Fig.\ref{fig:ene_diag}. For this simulation, a Lindblad equation is solved to find the time-dependent density matrix elements of the system,
\begin{eqnarray}
\label{Lindblad}
    \frac{d\rho(t)}{dt} = -\frac{i}{\hbar}[H(t), \rho(t)] + \mathcal{L}_D\rho(t) 
\end{eqnarray}
The Hamiltonian is given by $H(t) = \Omega + \bm{\mu}\cdot (\bm{E}_{\text{pump}}(t) + \bm{E}_{\text{probe}}(t+\tau))$ , where  $\Omega$ represents the field-free part of the Hamiltonian of the molecule and $\bm{\mu}$ denotes the dipole moments, all taken to be collinear along the molecular axis. The $\bm{E}_{\text{pump}}(t)$ and $\bm{E}_{\text{probe}}(t)$  correspond to pump and probe fields, respectively, and are modeled as linearly polarized Gaussian pulses, polarized along the same direction of the form $E(t) \propto \cos{(\omega_0 t + \phi)}\exp{(-t^2/2\sigma^2)}$. Each pulse has a photon energy of $4$ eV ($=\hbar\omega_0$) and a pulse width of $20$ fs. The pump pulse bandwidth is chosen to be sufficiently broad to populate all the excited states coherently. The decay and dephasing times contained in the Lindbladian $\mathcal{L_D}$ are set at $5$ ps and are sufficiently long compared to the simulation length of $1$ ps to not impact the coherent dynamics significantly. Eq~\ref{Lindblad} is solved numerically using the Euler method with a time step of $0.1$ fs as implemented in the UTPS code~\cite{utps2023}.

After solving for $\rho(t)$, the total time-dependent polarization can be calculated using $\bm{P}(t) = \text{Tr}[\bm{\mu \rho(t)}]$. However, to get the transient absorption output signal, we apply the phase-matching condition \cite{Haug2009},
\begin{align}
\label{phase_matching}
    \bm{P}_{\text{out}}(t) = \int d^2\bm{\phi} e^{i\bm{\phi}\cdot \bm{n}} \bm{P}(t, \bm{\phi})
\end{align}
where $\bm{\phi} = \{\phi_1, \phi_2\}$ are the phases of the pump and the probe and the vector $\bm{n} = \{0,1\}$ is chosen such that the signal is in the direction of the probe. To obtain the spectrum of the output signal, the Fourier transformation of $P_{\text{out}}(t)$ is performed. Finally, the average over all the angles between the molecular axis and the polarization of the probe/pump pulse is taken, using 200 angles sampled between $0$ and $\pi$.

\begin{figure*}
    \centering
    \includegraphics[width=0.8\linewidth]{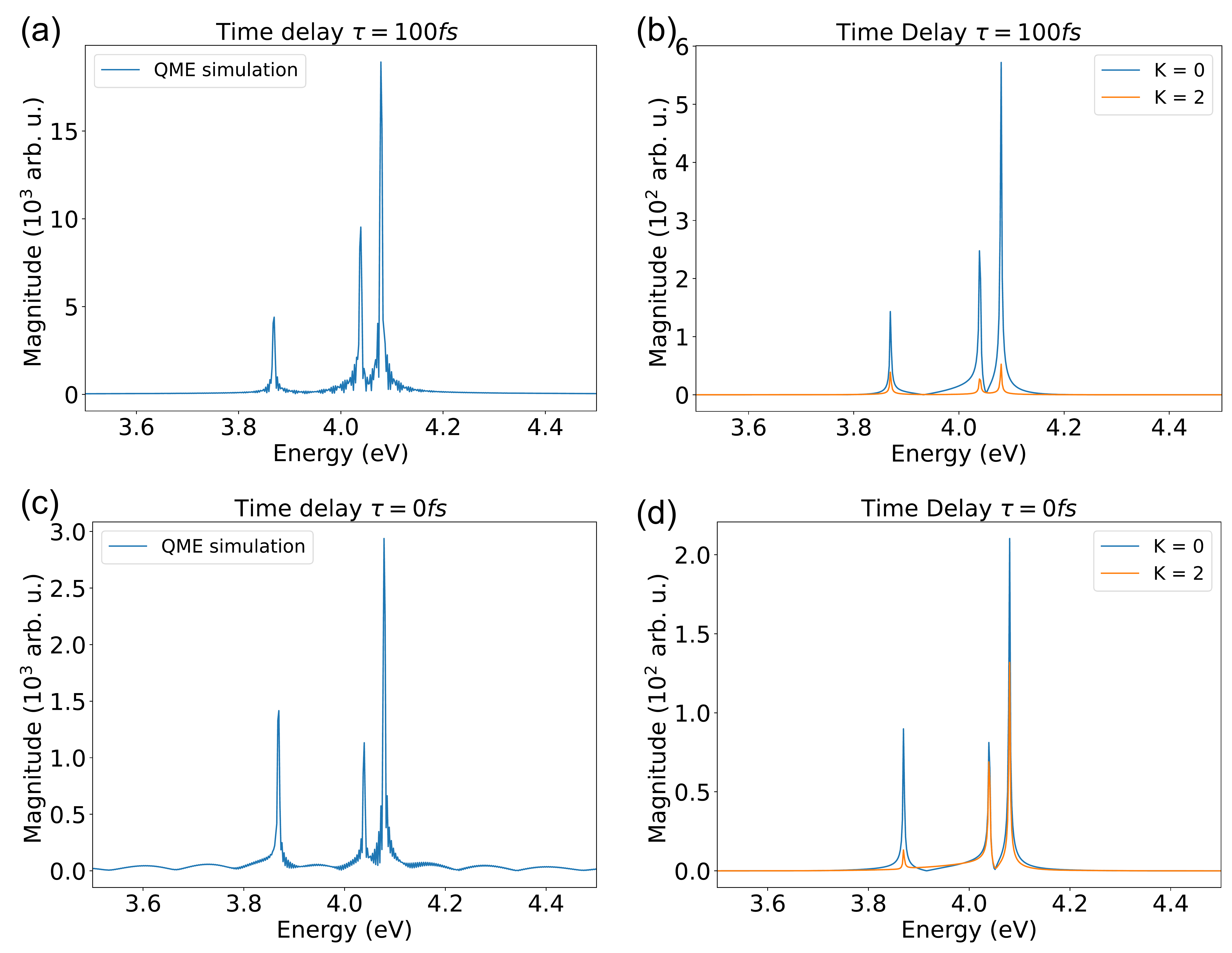}
    \caption{(a) and (c) show the polarization spectra of the transient absorption experiment described in Figure~\ref{fig:ene_diag} using the orientation-averaged QME simulation with $\tau = 100\text{ fs}$ and $\tau = 0\text{ fs}$ respectively. These spectra are compared with (b) and (d), which depict the LF first-order polarization spectra calculated using Equation~\ref{P1_MADM}. The former shows a good agreement with the latter's $K=0$ case, which corresponds only to the isotropic contribution. However, the higher-order moments also contribute to the LF polarization, capturing the anisotropic molecular axis distribution.}
    \label{fig: TDSE_comp_P1}
\end{figure*}

Fig~\ref{fig: TDSE_comp_P1}(a) and (c) present orientation-averaged first-order polarization obtained from the orientation-averaged QME simulation for time delays of 100 fs and 0 fs between the pump and probe pulses, respectively. Using identical input parameters, this simulation is compared with the first-order LF polarization derived in Eq.~\ref{P1_MADM}. Since the molecules are linear and the pulses are linearly polarized, the selection rules in Sec~\ref{subsubsection: Selection_rules} dictate that $Q=0, S=0$. Consequently, the only values allowed for the total angular momentum quantum number K are $K=0,2$. The first-order LF polarization spectrum for these two allowed K-values are plotted separately in Fig.~\ref{fig: TDSE_comp_P1}(b) for a time delay of 100 fs and in Fig.~\ref{fig: TDSE_comp_P1}(d) for a time delay of 0 fs between the pump and probe pulses (to calculate Eq~\ref{P1_MADM}, the terms corresponding to $K=0$ and $K=2$ must be added).

The orientation-averaged first-order polarization contains contributions only from $K=0$ terms, as the laboratory frame (LF) density matrices, when averaged over all orientations, are determined solely by isotropic MADMs \cite{Makhija2022}. This is evident from Eq~\ref{euler_basis}, where $K=Q=S=0$,
\begin{align}
    \int d\bm{\Omega} \rho_{n,n'}(\bm{\Omega}, \tau) =  8\pi^2 A_{00}^0(n, n'; \tau)
\end{align}

However, in a real experiment, orientation-dependent anisotropy is captured by higher-order total angular momentum terms ($K>0$).  As shown in Fig.~\ref{fig: TDSE_comp_P1}(b) and (d), the contributions for $K=2$ are non-zero and significant. This indicates that averaging over all orientations of the output polarization signal results in the loss of the anisotropic information. Comparison between the orientation-averaged QME simulation in Fig.~\ref{fig: TDSE_comp_P1}(a) and $K=0$ in Fig.~\ref{fig: TDSE_comp_P1}(b) for a time delay of 100 fs shows good agreement. However, the same does not hold for Fig~\ref{fig: TDSE_comp_P1}(c) and Fig.~\ref{fig: TDSE_comp_P1}(d), where the pulses are completely overlapped. This disagreement between the orientation-averaged QME simulation and the LF polarization is expected because, in deriving Eq~\ref{P1_MADM} in Sec~\ref{subsection: LF_P1}, it was assumed that the coherence or population created by the pump pulse remains unaffected by the probe pulse. This assumption breaks down when the pulses overlap as both the pump and probe pulses simultaneously populate the excited states.

\section{Towards Ultrafast Transient Quantum state Tomography using Transient Absorption Spectroscopy}
\label{Quantum_tomography}

\begin{figure}[ht]
    \centering
    \includegraphics[width=\linewidth]{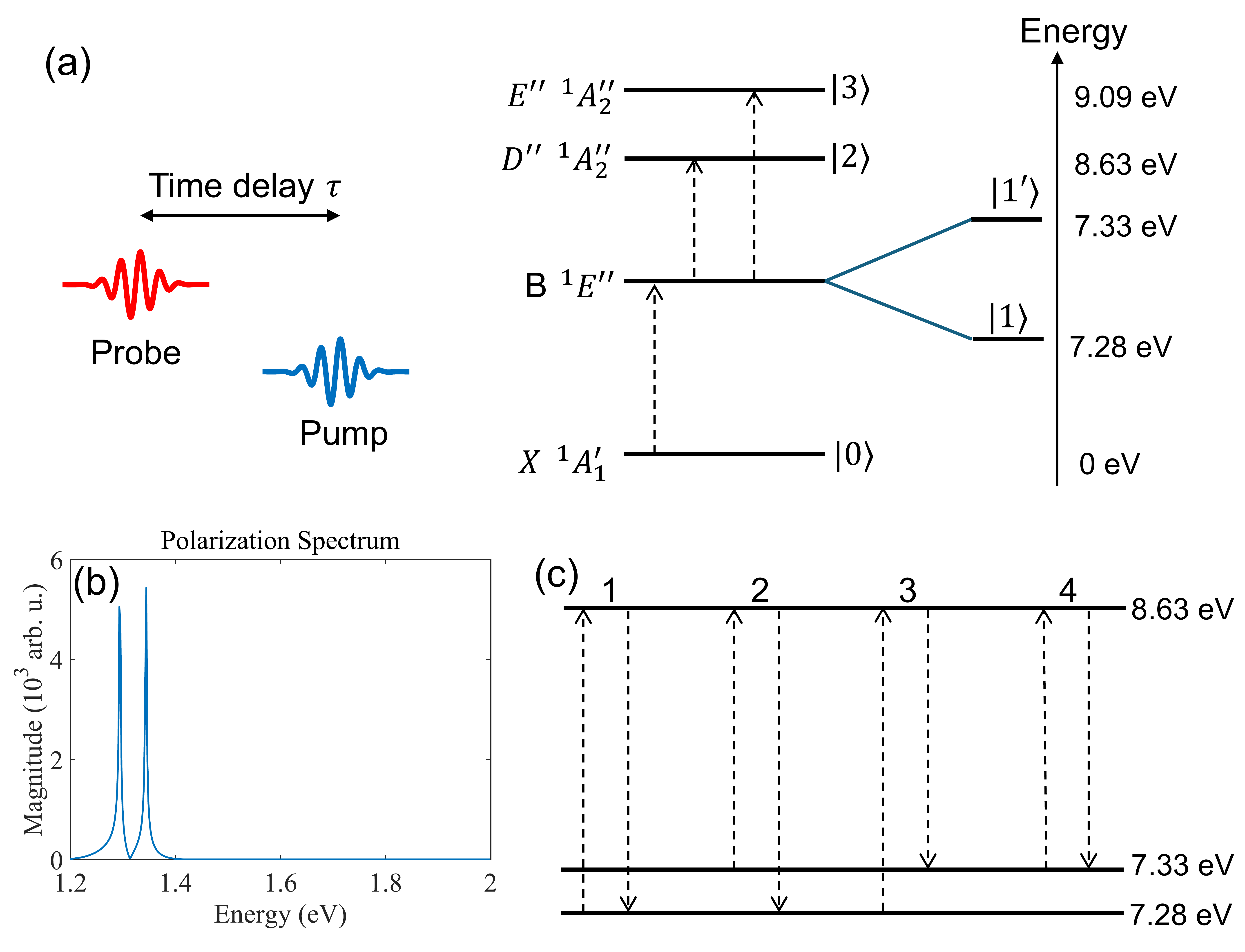}
    \caption{(a) The energy level diagram of ammonia molecule and the schematic of transient absorption spectroscopy. The pump has the photon energy of 7.3 eV to coherently populate $B\ ^{1}E''$ states. The probe pulse can have photon energy of either 1.3 eV or 1.8 eV for resonant excitation $D''\ ^{1}A_2'' \leftarrow B\ ^{1}E''$ or $E''\ ^{1}A_2'' \leftarrow B\ ^{1}E''$. (b) The example absorption spectrum for transition $D''\ ^{1}A_2'' \leftarrow B\ ^{1}E''$, where the peaks are centered at 1.295 eV and 1.34 eV. There are multiple pathways that contributes to the two peaks illustrated in (c).}
    \label{fig:ammonia_sch}
\end{figure}

Ultrafast Quantum Tomography involves the experimental reconstruction of time-dependent lab frame density matrix elements that fully characterize the electronic quantum state of a molecule. This has been achieved in a previous study \cite{PhysRevLett.131.193001} using photoelectron spectroscopy. The photoelctron angular distribution (PAD) anisotropy parameters are expanded in terms of MADMs. Because the parameters are measured using a time-resolved photoionization experiment, the time-dependent MADMs are extracted. The time-dependent LF density matrix can be fully reconstructed from the extracted MADMs using Eq~\ref{euler_basis}.

Here we propose that Quantum Tomography can also be achieved through optical observables using pump-probe transient absorption spectroscopy. To demonstrate this theoretically, we consider the ammonia molecule, whose electronic states are shown in Fig.\ref{fig:ammonia_sch} (a). These low-lying electronic states and their symmetries are well known \cite{Allen1991, Saraswathy2010}. Ammonia can be excited  resonantly using the pump pulse from the ground state $X\ ^{1}A_1'$ to a nearly degenerate state $B\ ^{1}E''$, $\lvert 1\rangle, \lvert 1'\rangle$, as shown in Figure~\ref{fig:ammonia_sch} (a). Previously, it has been shown that there are only three unique MADMs corresponding to the states $B\ ^{1}E''$ \cite{PhysRevLett.131.193001, Makhija_2020}.
\begin{subequations}
\label{ammonian_MADM}
\begin{align}
     A^0_{00}(1,1,t)=A^0_{00}(1',1',t)\\
     A^2_{00}(1,1,t)=A^2_{00}(1',1',t)\\
     A^2_{02}(1',1,t)=A^2_{0-2}(1,1',t)
\end{align}
\end{subequations}
As illustrated in the previous sections, $A^0_{00}(1,1,t)$ corresponds to the isotropic distribution of the population of the excited state. Whereas $A^2_{00}(1,1,t)$ signifies the anisotropy in the population of the state, specifically the alignment of the population with respect to the z-axis of the lab frame (pump polarization direction). The $A^2_{02}(1',1,t)$ tracks the orientation of the coherence between the two nearly degenerate states. To probe these states, we excite the $B\ ^{1}E''$ states resonantly to the higher-lying states $D''\ ^{1}A_2''$ or $E''\ ^{1}A_2''$, as shown in Fig.~\ref{fig:ammonia_sch} (a).

For the probe pulse with photon energy $1.3\ eV$, there will be a resonant transition $D''\ ^{1}A_2'' \leftarrow B\ ^{1}E''$ and the absorption spectrum contains two peaks (see Fig.~\ref{fig:ammonia_sch} (b)). The peaks in the absorption spectrum are the result of multiple pathways shown in Fig.~\ref{fig:ammonia_sch} (c). The pathways $1 + 2$ will contribute to the spectrum centered at $1.34eV$ and the pathways $3+4$ will count towards the spectrum at $1.295eV$. Eq.~\ref{P1_MADM} can be used to investigate which nonzero MADMs will contribute to the two peaks. For more details, see the supplementary material. However, to extract three MADMs that corresponds to the state $B^1E''$ in Eq.~\ref{ammonian_MADM}, we need at least one more spectral peak. This can be obtained by using a probe pulse of energy $1.8eV$ to resonantly excite to the state $E''\ ^{1}A_{2}''$. One can measure a similar spectrum as in Fig~\ref{fig:ammonia_sch} (b) and with similar pathways as Fig~\ref{fig:ammonia_sch} (c), which should be centered at $1.81eV$ and $1.76eV$. Now that we have enough equations, we can write Eq.~\ref{P1_MADM} in martix form as $P(\tau)=\mathcal{C}A(\tau)$. Since the matrix containing the coefficients $\mathcal{C}$ is not a square matrix, we can invert the matrix equation using the least square method as $A(\tau) = (\mathcal{C}^{T}\mathcal{C})^{-1}\mathcal{C}^{T}P(\tau)$ for each time delay $\tau$. From the extracted MADMs, the LF density matrix can be reconstructed thus achieving ultrafast time-resolved quantum state tomography.

\section{Field Observables}

Thus far, we have analyzed the spectrum of the polarization of the nonlinear process of transient absorption. In previous work, it has been shown that is possible go beyond measurement of intensity of light and measure the phase of the nonlinear signal's electric field \cite{sederberg2020, srivastava2024, kneller2025attosecond, walz2022, Pandey24}. If the first-order polarization is expressed as $P^{(1)}(\omega, \tau) = |P^{(1)}(\omega, \tau)|\exp{(i\phi_P(\omega, \tau)})$, then by measuring the phase of the electric field emitted by $P^{(1)}(\omega, \tau)$ it is possible to directly measure the phase $\phi_P(\omega, \tau)$. Utilizing Eq.\ref{P1_MADM}, Eq.\ref{euler_basis}, and the selection rules for linear molecules discussed in Sect.~\ref{subsubsection: Selection_rules}, the polarization can be written as,
\begin{eqnarray}
    P^{(1)}(\omega; \tau) = \sum_{\alpha, v_\alpha}\sum_{\beta, v_\beta} \sum_{K,Q,S}\int d\bm{\Omega}|\mathcal{K}(\omega)|\nonumber\\
    \times \exp{\left(i(\phi_{\mathcal{D}}^{\alpha\beta}(q)+\phi_E(\omega)+\phi_{\rho}^{\alpha\beta}(\bm{\Omega},\tau))\right)}
\end{eqnarray}

where, $\phi_{\mathcal{D}}^{\alpha\beta}(q)$ is the phase mixing of the MF dipole moments of the states $\alpha$ and $\beta$, $\phi_E(\omega)$ is the phase of the probe electric field, and $\phi_{\rho}^{\alpha\beta}(\bm{\Omega},\tau)$ corresponds to the phase of the density matrix elements between the states. The term $|\mathcal{K}(\omega)|$ contains all the amplitude contributions. Notably, when $\alpha = \beta$, the phase $\phi_{\rho}^{\alpha\alpha}(\bm{\Omega},\tau) = 0$, since the diagonal terms of the density matrix are real, since they correspond to the population of the state. Similarly, we find $\phi_{\mathcal{D}}^{\alpha\alpha}(q) = 0$, indicating that pathways that originate and terminate within the same state do not contribute to the phase of the output field. Instead, the variation in the phase of the output field arises from the phase of the coherence between two states.

The phase of the output field could be highly sensitive to the molecular orientation with respect to the incoming electric field \cite{Pandey24}. Thus the measurement of the phase along with the amplitude could provide additional, information on the phase evolution of molecular systems, independent from the populations, that can aide in performing Quantum Tomography. This will be explored in future work.

\section{Summary}
In conclusion, we have shown that electronic polarization corresponding to a pump-probe transient absorption signal can be written in terms of Molecular Axis Distribution Moments (MADMs) and coefficients that depend on molecular frame transition dipole moments. The MADMs contain the lab frame (LF) density matrix, which in most pump-probe experiments is not isotropic. When comparing experimental transient absorption signals from gas phase molecules to theoretical single molecule calculations, rotational averaging is typically performed. By comparing the calculated polarization to a Quantum Master Equation simulation of a transient absorption signal we have shown that rotational averaging of the calculated signal does not capture the anisotropic contributions due to higher order MADMs. Furthermore, we have described a path towards ultrafast quantum state tomography using transient absorption spectroscopy where measuring the transient absorption spectrum could provide access to the MADMs and hence to the time-resolved LF density matrix. Finally, we have discussed the potential of measuring electric field observables in nonlinear spectroscopy to further enhance quantum state tomography by gaining access to the phase of the polarization. Our work is applicable to a multitude of ultrafast nonlinear optical scenarios in gas-phase molecules and when combined with field-resolved spectroscopies could have important implications for quantum-state tomography in molecules on femtosecond and attosecond time scales.




\begin{acknowledgments}
This work was supported by the U.S. Department of Energy, Office of Science, Basic Energy Sciences, under Award $\#$ DE-SC0024234. Work at the Molecular Foundry was supported by the Office of Science, Office of Basic Energy Sciences, of the U.S. Department of Energy under Contract No. DE-AC02-05CH11231.
\end{acknowledgments}

\section*{Data Availability Statement}

The data that support the findings of this study are available from the corresponding authors upon reasonable request.

\bibliography{aipsamp}

\end{document}


\title{Supplementary material for the article: \\ Molecular Axis Distribution Moments in Ultrafast Transient Absorption Spectroscopy: A Path Towards Ultrafast Quantum State Tomography}



\author{Shashank Kumar}
    \email{kumar414@purdue.edu}
    \affiliation{Department of Physics and Astronomy, Purdue University, 525 Northwestern Ave., West Lafayette, 47907, Indiana, USA}%

\author{Eric Liu}%
    \affiliation{Department of Physics and Astronomy, Purdue University, 525 Northwestern Ave., West Lafayette, 47907, Indiana, USA}%

\author{Liang Z. Tan}
    \affiliation{Molecular Foundry, Lawrence Berkeley National Laboratory, Berkeley, California 94720, USA}

\author{Varun Makhija}
    \email{vmakhija@umw.edu}
    \affiliation{Department of Chemistry and Physics, University of Mary Washington, 1301 College Avenue, Fredericksburg, Virginia 22401, USA}

\author{Niranjan Shivaram}%
    \email{niranjan@purdue.edu}
    \affiliation{Department of Physics and Astronomy, Purdue University, 525 Northwestern Ave., West Lafayette, 47907, Indiana, USA}%
    \affiliation{Purdue Quantum Science and Engineering Institute, Purdue University, 1205 W State St, West
    Lafayette, 47907, Indiana, USA.}
\maketitle

\section{LF to MF dipole matrix elements}
\label{Appendix: LF_to_MF}
The LF dipole moment can be written in terms of its spherical components as\cite{Stolow2008, Makhija2022},
\begin{align}
\label{LF_sh}
    \bm{\mu} \cdot \hat{\epsilon} = \sum_{p=-1}^{1} (-1)^p d_p e_{-p}
\end{align}
and the spherical component can be rotated to the MF using the Euler angles $\bm{\Omega}=\{\phi,\theta,\chi\}$,
\begin{equation}
    d_p = \sum_{q=-1}^{1} D_{pq}^{1*}(\bm{\Omega}) d_q
\end{equation}
where $d_p,d_q$ are the dipole moment operators in the LF and MF respectively in the spherical component basis. $D^J_{KM}(\bm{\Omega})$ is the Wigner D-matrix and is defined for a symmetric top basis as\cite{Zare1988},
\begin{eqnarray}
\label{D_matrix}
    \langle \bm{\Omega}\lvert J K M \rangle = \sqrt{\frac{2J + 1}{8\pi^2}} D_{MK}^{J*}(\bm{\Omega})
\end{eqnarray}
Plugging all the definitions in the LF dipole matrix element,
\begin{align}
    &\langle n_\gamma M_\gamma \lvert \bm{\mu} \cdot \hat{\epsilon} \lvert n_\alpha M_\alpha \rangle \int d\bm{x} \langle n_\gamma M_\gamma \lvert \sum_{p=-1}^{1}\sum_{q=-1}^{1} (-1)^p  e_{-p} D_{pq}^{1*}(\bm{\Omega}) d_q \lvert \bm{x}\rangle \langle\bm{x} \lvert n_\alpha M_\alpha \rangle
\end{align}
In the last step, a new basis set $\lvert \bm{x}\rangle$ is introduced such that $\int \lvert \bm{x}\rangle \langle\bm{x} \lvert = 1$. The $\bm{x} = \{\bm{r}, \bm{R}, \bm{\Omega}\}$ is the combination of electronic coordinates, nuclear coordinates, and the Euler angles. Using the definition in Eq~\ref{D_matrix}
\begin{align}
    \langle n_\gamma M_\gamma \lvert \bm{\mu} \cdot \hat{\epsilon} \lvert n_\alpha M_\alpha \rangle = &\iint d\bm{r}\ d\bm{R}\sum_{p=-1}^{1}\sum_{q=-1}^{1} (-1)^p  e_{-p} \langle \bm{r},\bm{R}\lvert v_\alpha,\alpha\rangle d_q \langle v_\alpha, \alpha\lvert \bm{r},\bm{R}\rangle\nonumber\\
    &\times \int d\bm{\Omega} \frac{\sqrt{[J_\alpha, J_\gamma]}}{8\pi^2} D_{pq}^{1*}(\bm{\Omega}) D_{M_\alpha K_\alpha}^{J_\alpha *}(\bm{\Omega}) D_{M_\gamma K_\gamma}^{J_\gamma}(\bm{\Omega})
\end{align}
After, using the definition in Eq.3 of the main text, one can easily arrive at Eq.8.

\section{Derivation of Eq. 10}
\label{Appendix: derivation_3j}
Using the Equation (3.118) of Reference~\cite{Zare1988},
\begin{align}
    \int d\bm{\Omega} D_{pq}^{1*}(\bm{\Omega}) D_{M_\alpha K_\alpha}^{J_\alpha *}(\bm{\Omega}) D_{M_\gamma K_\gamma}^{J_\gamma}(\bm{\Omega})\notag &= \int d\bm{\Omega} (-1)^{p-q+M_\alpha - K_\alpha} D_{-p-q}^{1}(\bm{\Omega}) D_{-M_\alpha -K_\alpha}^{J_\alpha}(\bm{\Omega}) D_{M_\gamma K_\gamma}^{J_\gamma}(\bm{\Omega})\notag\\
    &= 8\pi^2 (-1)^{p-q+M_\alpha - K_\alpha} \begin{pmatrix}
        J_{\gamma} & J_\alpha & 1\\
        M_{\gamma} & -M_\alpha & -p
    \end{pmatrix}
    \begin{pmatrix}
        J_{\gamma} & J_\alpha & 1\\
        K_{\gamma} & -K_\alpha & -q
    \end{pmatrix}
\end{align}
Similarly,
\begin{align}
    \int d\bm{\Omega} D_{p'q'}^{1*}(\bm{\Omega}) D_{M_\beta K_\beta}^{J_\beta}(\bm{\Omega}) D_{M_\gamma K_\gamma}^{J_\gamma * }(\bm{\Omega})\notag &= \left\{\int d\bm{\Omega} D_{pq}^{1}(\bm{\Omega}) (-1)^{M_\beta - K_\beta} D_{-M_\beta -K_\beta}^{J_\beta}(\bm{\Omega}) D_{M_\gamma K_\gamma}^{J_\gamma}(\bm{\Omega})\right\}^*\notag\\
    &= 8\pi^2 (-1)^{M_\beta - K_\beta} \begin{pmatrix}
        J_{\gamma} & J_\beta & 1\\
        M_{\gamma} & -M_\beta & p'
    \end{pmatrix}
    \begin{pmatrix}
        J_{\gamma} & J_\beta & 1\\
        K_{\gamma} & -K_\beta & q'
    \end{pmatrix}
\end{align}

Using these relations and eliminating the Wigner-D matrices in Eq. 8, gives Eq. 10.

\section{First Order Polarization in Molecular Frame}
\label{Appendix: P1_MF}
As we derived the expression of the LF first-order polarization, we can follow a similar derivation starting from Eq. 7 of the main text. However, the total wavefunction in the molecular frame will be a linear combination of only vibronic states,
\begin{equation}
\label{MF_BO_basis}
    U(t)\lvert \Psi(0)\rangle = \lvert \Psi(t)\rangle = \sum_{\alpha, v_\alpha} C_{\alpha, v_\alpha}(t) e^{-i\mathcal{E}_{\alpha, v_\alpha} t} \lvert v_\alpha\rangle \lvert \alpha \rangle
\end{equation}
This means that the polarization spectrum now depends on the orientation of the molecular axis with the polarization of the interacting pulse. We introduce the Euler angle $\bm{\Omega}$ between the molecular axis and fixed Lab frame axis.
\begin{align}
\label{P1_MF}
    \Tilde{P}^{(1)}(\omega, \bm{\Omega}; \tau) = \sum_{\substack{\alpha, v_\alpha \\ \beta, v_\beta}}C_{\alpha, v_\alpha}(\tau) C_{\beta, v_\beta}^*(\tau) \times \sum_{\gamma, v_\gamma}\langle \beta, v_\beta\lvert \bm{\mu} \cdot \hat{\epsilon} (\bm{\Omega}) \lvert \gamma, v_\gamma \rangle \langle \gamma, v_\gamma \lvert \bm{\mu} \cdot \hat{\epsilon}(\bm{\Omega}) \lvert \alpha, v_\alpha \rangle \notag \\
    \times E'\left(\omega,\mathcal{E}_{\beta, v_\beta}, \mathcal{E}_{\alpha, v_\alpha}, \mathcal{E}_{\gamma, v_\gamma} \right)
\end{align}

 Similar to Eq.~\ref{LF_sh}, the dipole moment projection to the fixed Lab frame axis (in case of linear polarization along the direction of polarization axis) i.e. $\bm{\mu}\cdot \hat{\epsilon}$ can be written in terms of Molecular Frame (MF) spherical harmonics as,
\begin{align}
\label{MF_sh}
    \bm{\mu} \cdot \hat{\epsilon} = \sum_{q=-1}^{1} (-1)^{-q} d_{-q} e_{q}
\end{align}
and the MF spherical harmonics can be rotated to LF spherical harmonics,
\begin{align}
    e_{q} = \sum_{p=-1}^1 D_{pq}^1(\bm{\Omega})e_{-p}
\end{align}
Using the multiplication properties of Wigner-D matrices\cite{Zare1988}, the MF first-order polarization spectrum can be simplified to,
\begin{align}
\label{MF_P1}
    \Tilde{P}^{(1)}(\omega, \bm{\Omega}; \tau) = &\sum_{\substack{\alpha, v_\alpha \\ \beta, v_\beta}}\sum_{KQS} C_{KQS}^{MF}(\omega,\alpha, \beta, v_{\alpha},v_\beta) \times \frac{2K+1}{8\pi^2} \rho(\beta,\alpha; \tau) D^{K}_{QS}(\bm{\Omega})
\end{align}
where $C_{KQS}^{MF}(\omega,\alpha, \beta, v_{\alpha},v_\beta)$ is the MF coefficient and given by,
\begin{align}
    C_{KQS}^{MF}(\omega,\alpha, \beta, v_{\alpha},v_\beta) = \sum_{p',q'}\sum_{p,q}(-1)^{p+p'}8\pi^2e_{-p'}e_{-p} 
    \begin{pmatrix}
        1 & 1 & K\\
        -p' & -p & Q
    \end{pmatrix}
    \begin{pmatrix}
        1 & 1 & K\\
        -q' & -q & S
    \end{pmatrix}\notag\\
    \times \sum_{\gamma, v_\gamma}\mathcal{D}_{\beta\gamma}(-q')\mathcal{D}_{\gamma\alpha}(-q)E'\left(\omega,\mathcal{E}_{\beta, v_\beta}, \mathcal{E}_{\alpha, v_\alpha}, \mathcal{E}_{\gamma, v_\gamma} \right)
\end{align}

\section{Pathways contributing to the Absorption spectrum}
\label{Appendix: Pathways}
The absoption spectra peaks can be studied using Eq. 12. The equations corresponding to the transitions $B\ ^{1}E'' \leftarrow  D''\ ^{1}A_2'' \leftarrow B\ ^{1}E''$ and $B\ ^{1}E'' \leftarrow E''\ ^{1}A_2'' \leftarrow B\ ^{1}E''$ are
\begin{subequations}
\begin{align}
    P_{\text{1.34 eV}}(\omega, \tau) =\underbrace{C_{000}(\omega; 1,1)A^0_{00}(1,1,\tau) + C_{200}(\omega; 1,1)A^2_{00}(1,1,\tau)}_{\text{(Pathway 1)}} \notag\\
    + \underbrace{C_{202}(\omega; 1',1)A^2_{02}(1',1,\tau)}_{\text{(Pathway 2)}}\\
    P_{\text{1.295 eV}}(\omega, \tau) = \underbrace{C_{000}(\omega; 1',1')A^0_{00}(1',1',\tau) + C_{200}(\omega; 1',1')A^2_{00}(1',1',\tau)}_{\text{(Pathway 3)}} \notag\\
    + \underbrace{C_{20-2}(\omega; 1,1')A^2_{0-2}(1,1',\tau)}_{\text{(Pathway 4)}}\\
    P_{\text{1.81 eV}}(\omega, \tau) = \underbrace{C'_{000}(\omega; 1,1)A^0_{00}(1,1,\tau) + C'_{200}(\omega; 1,1)A^2_{00}(1,1,\tau)}_{\text{(Pathway 1)}} \notag\\
    + \underbrace{C'_{202}(\omega; 1',1)A^2_{02}(1',1,\tau)}_{\text{(Pathway 2)}}\\
    P_{\text{1.76 eV}}(\omega, \tau) = \underbrace{C'_{000}(\omega; 1',1')A^0_{00}(1',1',\tau) + C'_{200}(\omega; 1',1')A^2_{00}(1',1',\tau)}_{\text{(Pathway 3)}} \notag\\
    + \underbrace{C'_{20-2}(\omega; 1,1')A^2_{0-2}(1,1',\tau)}_{\text{(Pathway 4)}}
\end{align}
\end{subequations}

where, $C_{KQS}$ contains the intermediate state $\gamma = D''\ ^{1}A_{2}''$ and $C^{'}_{KQS}$ contains the intermediate state $\gamma = E''\ ^{1}A_{2}''$. Given that the absorption spectrum can be measured experimentally, we can use any three of the four equations to extract the dynamical MADMs. To do that, we can write the system of equations in matrix form by discreetly setting the $\omega$ domain as $\{\omega_1, \omega_2 \cdots \omega_N\}$,
\begin{align}
\begin{pmatrix}
        P_{\text{1.34 eV}}(\omega_1, \tau)\\
        P_{\text{1.295 eV}}(\omega_1, \tau)\\
        P_{\text{1.81 eV}}(\omega_1, \tau) \\
        \vdots\\
        P_{\text{1.34 eV}}(\omega_N, \tau)\\
        P_{\text{1.295 eV}}(\omega_N, \tau)\\
        P_{\text{1.81 eV}}(\omega_N, \tau)
    \end{pmatrix} &=
    \begin{pmatrix}
        C_{000}(\omega_1; 1,1) & C_{200}(\omega_1; 1,1) & C_{202}(\omega_1; 1',1) \\
        C_{000}(\omega_1; 1',1') & C_{200}(\omega_1; 1',1') & C_{20-2}(\omega_1; 1,1') \\
        C^{'}_{000}(\omega_1; 1,1) & C^{'}_{200}(\omega_1; 1,1) & C^{'}_{202}(\omega_1; 1',1) \\
        \vdots & \vdots & \vdots \\
        C_{000}(\omega_N; 1,1) & C_{200}(\omega_N; 1,1) & C_{202}(\omega_N; 1',1)\\
        C_{000}(\omega_N; 1',1') & C_{200}(\omega_N; 1',1') & C_{20-2}(\omega_N; 1,1')\\
        C^{'}_{000}(\omega_N; 1,1) & C^{'}_{200}(\omega_N; 1,1) & C^{'}_{202}(\omega_N; 1',1)
    \end{pmatrix}
    \begin{pmatrix}
        A^{0}_{00}(1,1,\tau)\\
        A^{2}_{00}(1,1,\tau)\\
        A^{2}_{02}(1',1,\tau)
    \end{pmatrix}\\
    \implies P(\tau) &= \mathcal{C}A(\tau)
\end{align}

We can use the least-squares method to solve MADMs as $A(\tau) = (\mathcal{C}^{T}\mathcal{C})^{-1}\mathcal{C}^{T}P(\tau)$ for each $\tau$ independently.
\bibliography{aipsamp}